\documentclass{ws-ijmpa}
\newcommand{\be}{\begin{equation}}
\newcommand{\ee}{\end{equation}}
\newcommand{\ba}{\begin{eqnarray}}
\newcommand{\ea}{\end{eqnarray}}

\begin{document}

\markboth{G. Calder\'on and G. L\'opez Castro}
{Convolution formula and finite W boson width effects in top quark width}

%
\catchline{}{}{}{}{}
%

\title{Convolution formula and finite W boson width effects in top quark width}

\author{G. Calder\'on}

\address{Facultad de Ingenier\'ia Mec\'anica y El\'ectrica,
Universidad Aut\'onoma de Coahuila, C.P. 27000, Torre\'on, Coahuila, M\'exico\\
gecalde@gmail.com}

\author{G. L\'opez Castro\footnote{On leave absence from: Departamento de 
F\'isica, Cinvestav, A.P. 14-740, 07000  M\'exico D.F, M\'exico}}

\address{Instituto de F\'isica, Universidad Nacional Aut\'onoma de M\'exico,
A.P. 20-364, 04510 Mexico D.F., Mexico\\
glopez@fis.cinvestav.mx}

\maketitle

\begin{history}
\received{Day Month Year}
\revised{Day Month Year}
\end{history}

\begin{abstract}

In the standard model, the top quark decay width $\Gamma_t$ is computed from the 
exclusive $t\to bW$ decay. We argue in favor of using the three body decays 
$t\to bf_i\bar{f}_j$ to compute $\Gamma_t$ as a sum over these exclusive modes. As 
dictated by the S-matrix theory, these three body decays of the top quark involve 
only asymptotic states and incorporate the width of the $W$ boson resonance in a 
natural way.  The convolution formula (CF) commonly used to include the finite
width effects is found to be valid, in the general case, when the intermediate 
resonance couples to a conserved current (limit of massless fermions in the case of 
$W$ bosons). The relation $\Gamma_t=\Gamma(t\to bW)$ is recovered by taking the limit 
of massless fermions followed by the $W$ boson narrow width approximation. Although both 
calculations of $\Gamma_t$ are different at the formal level, their results would differ 
only by tiny effects induced by light fermion masses and higher order radiative corrections.

\end{abstract}

\ccode{PACS numbers: 11.10.St, 13.38.Be, 14.65.Ha, 14.70.Fm}

\section {Introduction}

The precise calculation and measurement of the top quark decay width $\Gamma_t$ is important 
to provide a consistency check of the Standard Model because $\Gamma_t$ depends on the 
cubic of the top quark mass. Moreover, the large mass of the top quark implies a large
decay width which in turn indicate that the top cannot be bounded to form hadrons \cite{orr,orr-1}. 
This feature makes attractive the application of perturbative methods to evaluate with high 
accuracy the quantum corrections to top quark decays. In particular, one expects that 
the top quark decay would provide the QCD analog of muon decay at the level of radiative
corrections and that useful information about $\alpha_s(m_t)$ can be extracted. Therefore a 
precise definition of $\Gamma_t$ incorporating all identified sources of corrections is 
necessary.

As is usually done, theoretical predictions have to match an accuracy comparable to or better 
than the experimental error bars. On the experimental side, the mass of the top quark $m_t$
was expected to be measured with an uncertainty of up to 3 GeV in the Run II at the Tevatron  
Collider \cite{willenbrock,willenbrock-1,willenbrock-2}, but actually the experimental results have improved the 
expectations and we have, including new preliminary Run II results from CDF and D0, an 
uncertainty of 1.4 GeV (see Ref.~ \refcite{tevatron}), while $\Gamma_t$ may reach a precision 
of $5-7\%$ from the forward-backward asymmetry at planned linear colliders \cite{fujii97}. 
On the theoretical side, it is customary to compute the top quark width from the $t\to bW$ 
decay \footnote{The decay modes $t\to sW$, $dW$ would be important only for a calculation 
aiming an accuracy below the $0.1\%$ level.} (see p. 517 in Ref.~ \refcite{pdg2006} and 
Refs.~ \refcite{jk,jk-1,alpha_s2,alpha_s2-1,alpha,alpha-1}). 
This is called the narrow 
width approximation because the W boson is considered to be a stable on-shell particle. 
Different radiative corrections to this process have been reported within the Standard Model. 
The order $\alpha_s$ and $\alpha_s^2$ QCD corrections to this rate turn out to be at the $10\%$ 
\cite{jk,jk-1} and $1\sim 2 \%$ \cite{alpha_s2,alpha_s2-1} level, respectively. On the other hand, the one 
loop electroweak corrections are found to affect the decay rate at the $1\sim 2\%$ level 
\cite{alpha,alpha-1}. Even some tiny effects in $\Gamma_t$ of ${\cal O}(10^{-5})$ arising from the 
renormalization of the $V_{tb}$ matrix element, have been reported in the literature
\cite{renorm}. Finally, given the large mass of the top quark, virtual effects of hypothetical 
heavy particles might contribute to $\Gamma(t\to bW)$ at a few of percent level 
(see for example Ref.~ \refcite{higgs, higgs-1}).

Another important correction to be applied to $\Gamma_t$ concerns the finite width effects 
of the $W$ boson ($\Gamma_W \not =0$). This was first considered in Ref.~ \refcite{jk,jk-1} within 
the factorization approximation and using the convolution formula. 
In this paper we approach this problem by looking at the transition amplitude 
for $t\to bf_if_j$ decays which considers the  finite propagation of the unstable $W$ 
boson and the finite fermion masses of its decay products. We also discuss the conditions 
that allows to derive the convolution formula to account for particle unstability and we 
comment on possible departures from it when higher order corrections are considered. 
Such an approach was implicitly used in Ref.~ \refcite{Soni} where it was found that the 
finite propagation of an unstable $W$ boson provides a source for the strong phase necessary 
to generate a non zero CP rate asymmetry in $t\to b\tau^\pm \nu_\tau$ decays. Clearly, this 
observable would not occur if the $W$ boson is considered to be a stable particle.

Since the original submission of this paper, other works have appeared which considers the 
finite width effects in the observables associated to the production and decay of top quarks 
and massive gauge bosons at hadron and $e^+e^-$ colliders \cite{Bernreuther,Bernreuther-1,Bernreuther-2}. 
In the present version of our paper we focus on effects of unstability of particles in the precision 
definition of their decay widths.

\section {Finite $W$ boson width effects}

As it was indicated above, most of the calculations of $\Gamma_t$ (with the exception of 
Refs.~ \refcite{jk,jk-1}) assume that the $W$ is a stable particle. The analysis of 
Refs.~ \refcite{jk,jk-1} 
shows that the finite width of the $W$ boson can induce an additional correction of
$1\sim2\%$ to the $t\to bW$ decay rate. In practice, the $W$ boson is not a real particle 
that can be reconstructed from their decay products with a well defined invariant mass $m_W$. 
Instead, the $W$ boson is a resonance and  it can not formally be considered as an asymptotic 
state to be used in the evaluation of S-matrix amplitudes. Furthermore, by cutting the 
production and decay mechanisms of a resonance can lead to miss important interference effects 
in the real observables which necessarily are related to the {\it detection} of (quasi)stable 
particles \cite{kz}.

For purposes of this paper we start by defining the top quark width
by using the three body decays $t\to bf_i\bar{f}_j$ as follows

\be \Gamma_t \equiv \sum \Gamma(t\to bf_i\bar{f}_j)\ . \label{decay}\ee

The sum is carried over flavors and colors of the fermion pair $f_i\bar{f}_j$ that are 
allowed by kinematics. Since the lifetimes of fermions in the final states are much larger 
than the typical interaction time scales (the $W$ and $t$ lifetimes), they can be considered 
as asymptotic states of this process. Note that if the production and decay of the 
intermediate $W$ boson are considered as independent processes, the above formula can be 
approximated by

\ba \Gamma_t &\approx& \Gamma(t\to bW)\times \sum_{i,j}
\frac{\Gamma(W\to f_i\bar{f}_j)}{\Gamma_W} = \Gamma(t \to bW)\ , \ea

\noindent which corresponds to the usual approximation.

We show in this paper that, in the limit of massless fermions, the tree level expression of 
Eq. (\ref{decay}) gives rise to a convolution formula commonly used (see for example 
Refs.~ \refcite{muta,tbwz}) to  include the finite width effects of the $W$ boson. 
Later, we let 
the $W$ boson width to vanish and we show that the r.h.s of Eq. (\ref{decay}) reduces to 
$\Gamma(t\to bW)$. 

Let us start our evaluation of the top decay width by computing the $t\to bf_i\bar{f}_j$ 
decay rate. The tree level amplitude for this process is given by

\ba {\cal M} &=& \frac{g}{2\sqrt{2}} V_{tb}\bar u(p_b)\gamma^\mu (1-\gamma_5)u(p_t) 
(-iD_{\mu\nu}(Q))
\frac{g}{2\sqrt{2}}V_{ij}^*\bar v(p_j)\gamma^\nu(1-\gamma_5)u(p_i)\ . \label{treelevel} \ea
 
In this expression $V_{kl}$ denote the corresponding Kobayashi-Maskawa matrix element 
associated to the $kl$ fermionic current ($V_{ij}=1$ for lepton doublets), $g$ is the 
strength of the weak charged current and the momentum transfer is defined by 
$Q=p_t-p_b=p_i+p_j$. In the unitary gauge, the resonant $W$ boson propagator $D_{\mu\nu}(Q)$ 
\cite{wprop1,wprop1-1} can be divided into its spin-1 (transverse) and spin-0 (longitudinal) pieces
according to

\ba D_{\mu\nu}(Q) &=& \frac{g_{\mu\nu}-\frac{\textstyle Q_{\mu}Q_{\nu}}{\textstyle Q^2}}
{Q^2-m_W^2+i Im\Pi_T(Q^2)}-
\frac{Q_{\mu}Q_{\nu}}{Q^2} \frac{1}{m_W^2-iIm\Pi_L(Q^2)}\ .\label{propaW} \ea

We consider $m_W$ to be the renormalized mass of the $W$ boson \footnote{Since we 
concentrate only on the imaginary parts of the $1PI$ corrections, the specific mass 
renormalization scheme is not of much relevance. However, this expression can be 
accommodated in the on-shell renormalization scheme.} and we include only 
(finite) absorptive corrections to the $W$ propagator as done in the context of the 
fermion loop scheme \cite{wprop1,wprop1-1,fls,fls-1}. In Eq. (\ref{propaW}) we have defined $Im \Pi_T$ and 
$Im\Pi_L$ as the transverse and longitudinal projections of the $W$ boson self energy,  
namely \cite{wprop1,wprop1-1}

\ba Im \Pi^{\alpha\beta}(Q) &=& \left(g^{\alpha\beta}-
\frac{Q^{\alpha}Q^{\beta}}{Q^2}\right) Im\Pi_T(Q^2)
+ \frac{Q^{\alpha}Q^{\beta}}{Q^2}Im \Pi_L(Q^2)\ . \ea

Using cutting rules techniques, we can compute these absorptive parts of the $W$ boson 
self energy. The expressions obtained at the lowest order in $g$ are given by

\ba Im\Pi_T(Q^2) &=& \sqrt{Q^2} \sum_{k,l} \Gamma^0(W(Q^2)
\rightarrow f_k\bar{f}_l) \theta(Q^2-(m_k+m_l)^2)\nonumber \\
Im\Pi_L(Q^2) &=& -\sum_{k,l}N_C \frac{g^2}{8}|V_{kl}|^2
\frac{Q^2}{4\pi} f(x_k^2,x_l^2)\lambda^{1/2}(1,x_k^2,x_l^2)
\theta(Q^2-(m_k+m_l)^2)\ , \label{absorW}\ea

\noindent where the sum extends over flavors ($k=u,c,e,\mu,\tau;\ l=s,d,b,
\nu_e,\nu_{\mu}, \nu_{\tau}$) running in fermion loops for a $W$ boson decay of squared 
momentum $\sqrt{Q^2} > m_k+m_l$. Let us emphasize that we have kept finite the masses of 
all fermions in these corrections in order to remain consistent when the masses of fermions 
in the final states of top decay are finite.

In the above expressions we have defined the (tree level) partial decay width of off shell 
$W$ bosons as follows

\ba \Gamma^0(W(Q^2)\to f_k\bar{f}_l) &=& N_C\left(
\frac{g^2}{8} \right) \frac{|V_{kl}|^2}{12\pi}\sqrt{Q^2} \left[
2-f(x_k^2,x_l^2)\right] \lambda^{1/2}(1,x_k^2,x_l^2)\ , \ea

\noindent where $N_C$ is the number of colors, $x_i\equiv m_i/\sqrt{Q^2}$ and we have defined 
the functions: $f(r,s)\equiv r+s+(r-s)^2$ and $\lambda(a,b,c)\equiv a^2+b^2+c^2-2(ab+ac+bc)$.

As we mention in the introduction, the top quark decay width and their radiative corrections 
reported in the literature are usually calculated by using the two body $t\to bW$ process.
If the $W$ boson is off its mass shell, with a mass $\sqrt{Q^2}$, the decay width of 
the top quark becomes

\ba \Gamma^0(&t& \rightarrow bW(Q^2)) \nonumber\\
&=& \frac{G_F^0 m_t^3}{8\pi\sqrt{2}}
|V_{tb}|^2 \frac{m_W^2}{Q^2}\left[(1-x^2)(1+2x^2)-y^2(2-x^2-y^2) \right]
\lambda^{1/2}(1,x^2,y^2)\ \label{top},\ea

\noindent where $x\equiv \sqrt{Q^2}/m_t$ and $y\equiv m_b/m_t$. Note that we have replaced 
$g^2\to 8G_F^0m_W^2/\sqrt{2}$, where $G_F^0$ is the Fermi constant at the tree level.

A straightforward calculation of the top quark decay width using Eq. (\ref{decay}) leads to 
the following result

\ba \Gamma_t^0 &=& \sum_{i,j}\Gamma^0(t\to bf_i\bar{f}_j)\nonumber\\
&=&\frac{1}{\pi}\sum_{i,j} \int_{(m_i+m_j)^2}^{(m_t-m_b)^2} dQ^2 \
\Gamma^0(t \to bW(Q^2)) \frac{\sqrt{Q^2}\ \Gamma^0(W(Q^2)
\to f_i\bar{f}_j)}{(Q^2-m^2_W)^2+(Im\Pi_T(Q^2))^2}\nonumber \\
&& +\ \frac{G_F^2 m_W^4 m_t^3}{64\pi^3}
\sum_{i,j}|V_{tb}V_{ij}^*|^2\int_{(m_i+m_j)^2}^{(m_t-m_b)^2} dQ^2
\lambda^{1/2}(1,x^2,y^2)\lambda^{1/2}(1,x_i^2,x_j^2) \nonumber \\
&& \ \ \ \ \ \ \ \ \times \left[(1-y^2)^2-x^2(1+y^2)\right]
\frac{f(x_i^2,x_j^2)}{m_W^4+(Im\Pi_L(Q^2))^2}\ . \label{prime}\ea

The first term in Eq. (\ref{prime}) arises from the spin-1 degrees of freedom in the $W$ 
boson propagator and the second one comes from its spin-0 component. The interference 
term in Eq. (\ref{prime}) vanishes due to the orthogonality of the amplitudes arising from 
the decomposition in Eqs. (\ref{treelevel},\ref{propaW}). Observe that the second term
in the r.h.s of Eq. (\ref{prime}) vanishes in the limit of massless fermions 
($f(x_i=0,x_j=0)=0$); if we keep the finite masses of these fermions, the second term in 
Eq. (\ref{prime}) will give a negligible contribution ($\sim 2.7\times 10^{-5} $ GeV) to 
$\Gamma_t$. The important point to stress here is that such corrections  are very small 
because the intermediate $W$ boson is coupled to an almost conserved vector current in
the final state. 

\section {Narrow width approximation}

Let us now consider the massless limit for  fermions that participate in the $W$ 
boson self energy correction and the decay of the $W$ boson.  In this case we have, from 
Eqs. (\ref{absorW}), $Im\Pi_L(Q^2)=0$ and

\ba Im\Pi_T(Q^2)&=& \sum_{k,l}\sqrt{Q^2}\Gamma(W(Q^2)\to f_k\bar{f}_l)
= \frac{Q^2}{m_W}\Gamma^0_W\ , \ea

\noindent where $\Gamma^0_W=\sum_{k_l}N_Cg^2m_W/48\pi$ is the on shell decay width of the $W$ 
boson in the limit of massless fermions. Thus, in the limit of massless (light) 
fermions, Eq. (\ref{prime}) is

\ba \Gamma^0_t &=& \int_0^{(m_t-m_b)^2} dQ^2\ \Gamma^0(t\to bW(Q^2))\rho_W(Q^2)\ ,
\label{convo}\ea 

\noindent where  

\ba \rho_W(Q^2) &=& \frac{1}{\pi} \frac{\frac{\textstyle Q^2}{\textstyle m_W}
\Gamma_W^0}{(Q^2-m_W^2)^2+\left (\frac{\textstyle Q^2}{\textstyle m_W}
\Gamma_W^0 \right)^2}\ . \label{density}\ea

The convolution kernel $\rho_W(Q^2)$ in Eq. (\ref{convo}) coincides with the one used in 
Refs.~ \refcite{muta,tbwz} to include the finite width effects of gauge bosons in final states. 
The factor $Q^2$ in the numerator of the convolution kernel serves to cancel the $Q^2$
factor appearing in the denominator of the $t\to bW(Q^2)$ decay rate -see Eq. (\ref{top})- 
and avoids that the integrand in Eq. (\ref{convo}) diverges in the limit $Q^2\to 0$. This 
result is consistent with the fact that in the limit $Q^2\to 0$, the $W$ boson produced in 
the $t\to bW(Q^2)$ decay has only two degrees of freedom and some care must be taking when 
using Eq. (\ref{top}) in that limit. On the other hand, Eq. (\ref{convo}) (and, as a matter 
of fact, the results of Refs.~ \refcite{muta,tbwz}) can be viewed as a factorization of the 
production and decay subprocesses of the $W$ gauge-boson. As already explained, this 
approximation (which can be justified on probabilistic grounds when the production and 
decay mechanisms of the $W$ boson are independent) can be valid only in the limit of 
massless fermions.

Next we focus in the narrow $W$ boson width approximation. In the limit $\Gamma_W^0\to 0$, 
Eq. (\ref{convo}) reduces to

\ba \Gamma_t^0&\to & \int dQ^2\Gamma^0(t \to bW(Q^2))\delta(Q^2-m_W^2) 
= \Gamma^0(t\to bW) \label{usual}\ .\ea

Thus, we consistently recover the tree level decay rate of  two body decay, 
Eq. (\ref{top}), which is used as the starting point to implement the radiative 
corrections to $\Gamma_t$ in other calculations \cite{jk,jk-1}\cdash\cite{higgs,higgs-1}.

One may wonder also which are the possible effects of considering the finite width of the 
top quark on its $t \to bW$  decay width. This question may be of relevance in the case 
that the top quark width is not extracted from a Breit-Wigner invariant mass distribution, 
as commonly done with other resonance parameters. An estimate of such effects can be done by 
using similar arguments as above, namely

\ba \Gamma^{smear}_t &=& \int^{(m_t+\Delta)^2}_{(m_t-\Delta)^2}dQ'^2\Gamma^0(t(Q'^2)\to
bW)\rho_t(Q'^2)\ , \ea

\noindent with a definition of $\rho_t(Q'^2)$ similar to Eq. (\ref{density}), with 
$(m_W,\ \Gamma_W)$ replaced by $(m_t,\ \Gamma_t)$.  If we choose $m_t=174.2$ GeV, 
$m_W=80.403$ GeV \cite{pdg2006} and $\Delta=20\Gamma^0_t(t\to bW)$, where $\Gamma^0_t(t\to
bW)$ is given by Eq. (8) when $Q^2=m_W^2$, we get

\ba \frac{\Gamma^{smear}_t}{\Gamma_t^0(t\to bW)}\approx 0.989 \ , \ea

\noindent namely it reduces the top quark decay width by $1.1\%$. We note that this effect 
is sensitive to the specific value used for $\Delta$.

To end this section, we provide the values of the top quark decay width that are calculated 
using the different approximations. For comparative purposes we normalize the different 
results to $\Gamma^0_t(t\to bW)$. Thus, we get

\ba \frac{\Gamma_t}{\Gamma_t^0(t\to bW)} &=& \left\{\begin{array}{c} 
\!\!\! 0.98508,\ \ \  \mbox{\rm using Eq. (\ref{prime})} \\
0.98509, \ \ \ \mbox{\rm using Eqs. (\ref{convo}-\ref{density})}\end{array} \right. \ .\ea

We observe that the effect of finite $W$ boson width is about $1.5\%$. Also, a comparison 
of the two numbers given above shows that finite fermion mass effects are negligible, of order 
$1\times 10^{-5}$, and that the CF of Eqs. (\ref{convo}-\ref{density}) gives an excellent 
approximation. In the above calculations we use the CKM parameters expressed in terms of 
the Wolfenstein parameters, whose central values we take $\lambda=0.2272$, $A=0.818$, 
$\bar \rho=0.221$ and $\bar \eta=0.340$ \cite{pdg2006}. We use for the Fermi coupling 
constant the value $G_F=1.16637 \times 10^{-5}$ $GeV^{-2}$ and for the CKM matrix element 
$V_{tb}=1$. For the lepton and quark masses we use values quoted in Ref.~ \refcite{pdg2006}.

\section{Departures from the convolution formula: an example}

We mention above that departures from the convolution formula (CF), Eq. (\ref{convo}), 
are expected when the intermediate resonance is not coupled to a conserved current.
In the example considered above, the $W$ boson couples to a quasi conserved current,
although the effects of finite fermion masses are very small. In this section we consider
another example where departures from the CF arising from a quasi conserved current can be
more sizable.

Let us consider the decay $\tau^-\to \nu_\tau V^{*-}(Q^2)\to \nu_\tau P_1 P_2$,  where
$P_1P_2=\pi^-\pi^0\ (K\pi)$ are pseudoscalar mesons and $V^{*-}=\rho^-(\ K^{*-})$ denotes an
intermediate vector meson. We choose the following propagator for the intermediate virtual
meson \cite{Genaro}

\be D^{\mu\nu}_V(Q^2) = \frac{g^{\mu\nu}  -\frac{\textstyle Q^{\mu}Q^{\nu}}{\textstyle
M_V^2-i\sqrt{Q^2}\Gamma_V(Q^2)}}{Q^2-M_V^2+i\sqrt{Q^2}\Gamma_V(Q^2)}\ , \ee

\noindent where $\Gamma_V(Q^2)$ denotes the energy dependent width of the vector meson and 
$M_V$ its mass.  As in the previous example, we decompose $D^{\mu\nu}_V(Q^2)$ into their 
transverse (T) and longitudinal (L) pieces; thus we get the following expression for the 
decay rate

\be \Gamma(\tau\to \nu_\tau P_1P_2) = \Gamma_T + \Gamma_L \ , \ee

\noindent where ($q=d$ or $s$)

\ba \Gamma_T &=& \int_{(m_1+m_2)^2}^{m_{\tau}^2}dQ^2 \frac{\sqrt{Q^2}
\Gamma(\tau \to V(Q^2)\nu_{\tau})\cdot \Gamma(V(Q^2)\to
P_1P_2)}{(Q^2-M_V^2)^2+Q^2\Gamma^2_V(Q^2)}\\  
\Gamma_L &=& \frac{G_F^2|V_{uq}|^2g_V^2g_{VP_1P_2}^2}{256\pi^3m_{\tau}}
\int_{(m_1+m_2)^2}^{m_{\tau}^2} \frac{dQ^2}{Q^6}
\frac{(m_{\tau}^2-Q^2)^2(m^2_1-m_2^2)^2  
\lambda^{1/2}(Q^2,m_1^2,m_2^2)}{M_V^4+Q^2\Gamma^2_V(Q^2)}\ .\ea 

In the above expression $g_V$ ($g_{VP_1P_2}$) denote the weak (strong) couplings of the
intermediate vector meson, and we have defined

\ba \Gamma(\tau \to V(Q^2)\nu_{\tau}) &=&\frac{G_F^2|V_{uq}|^2g_V^2}{16\pi m_{\tau}^3}
\cdot \frac{(m_{\tau}^2-Q^2)^2(m_{\tau}^2+2Q^2)}{Q^2} \\
\Gamma(V(Q^2)\to P_1P_2) &=& \frac{g_{VP_1P_2}^2}{48\pi}\cdot
\frac{\lambda^{3/2}(Q^2,m_1^2,m_2^2)}{(Q^2)^{5/2}}\ . \ea

Using the values $g_{\rho^-\pi^-\pi^0}= 5.959$ and  
$\sqrt{2}|g_{K^{*-}K^-\pi^0}|=|g_{K^{*-}\bar{K}^0\pi^-}|=4.575$, we get (note that the values 
of $g_V,\ G_F$ and $|V_{ij}|$ cancels in the ratio)

\ba \frac{\Gamma_T(\tau^- \to P_1P_2\nu_{\tau})}{\Gamma(\tau^- \to V^-\nu_{\tau})}
&=& \left\{ \begin{array}{c}
0.8615 \ \ \ \mbox{\rm for $\pi^-\pi^0$} \\ 
\!\!\!\! 0.9263 \ \ \ \mbox{\rm for $K\pi$} \end{array} \right. \\ 
\frac{\Gamma_L(\tau^- \to P_1P_2\nu_{\tau})}{\Gamma(\tau^- \to V^-\nu_{\tau})}
&=& \left\{ \begin{array}{c}
4.44\times 10^{-6} \ \ \ \mbox{\rm for $\pi^-\pi^0$} \\
\!\!\!\! 8.50 \times 10^{-3} \ \ \ \mbox{\rm for $K\pi$} \end{array} \right.\ , \ea

\noindent where  $\Gamma(\tau^-\to V^-\nu_{\tau})$ denotes the decay rate of Eq. (21) 
taken at $Q^2=M_V^2$ and $K\pi$ denotes the sum of $K^-\pi^0$ and $\bar{K}^0\pi^-$ 
channels. Thus, we observe that departures from the CF induced by the non conservation of 
the weak current in the hadronic vertex are small, and the most important effect 
(of ${\cal O}(1 \%)$) corresponds to the $K\pi$ channel (SU(3) breaking in $K\pi$ vs. SU(2)
breaking in $\pi\pi$). 
Also, we note that the finite width effects are more important as the  intermediate resonance 
is broader. 

\section {Conclusions}

In summary, the purpose of this paper is to point out that the top quark decay width 
must be evaluated from its three body decays ($t\to bf_i\bar{f}_j$)  instead of using 
the decay $t\to bW$. The former modes involve only final state particles that can be 
considered as asymptotic states required to correctly define S-matrix amplitudes. On the 
other hand, the three body decay amplitudes incorporate the finite width effects of the 
$W$ boson in a natural way. 

We find that the convolution formula (CF) commonly used to incorporate the finite width 
effects is valid when the intermediate $W$ boson is coupled to a conserved current (the 
spin-0 component of the $W$ boson propagator decouples in the limit of massless fermions). 
Clearly, the CF gives an excellent approximation to incorporate finite width effects in the 
top quark decay width at the tree level but it can not be the case in general. As an example 
we have considered $\tau^-\to P_1P_2\nu_{\tau}$ decays where departures from the CF at the 
level of $1\%$ can be obtained.

As we pointed out before, the non conservation of the weak charged current signal departures 
from the exact convolution formula for the top quark decay width. Note also that higher order 
corrections can also break the validity of the exact convolution formula. In particular, the 
box diagram electroweak corrections which connects the $tb$ to the $f_i\bar{f}_j$ external 
lines does not allows for a unique transverse plus longitudinal decomposition of the 
intermediate state ($WZ,\ W\gamma$) that connects these currents. Thus, additional 
contributions to the CF can be expected in the presence of higher order corrections. Note 
however that the one loop QCD corrections to $t \to bf_i\bar{f}_j$ will not affect the
convolution formula (Eq. (\ref{convo})) as long as the QCD box corrections and the
tree level amplitudes do not interfere at this order \cite{fjy,smith}.

\section*{Acknowledgments}

We are grateful to J. Pestieau, F. J. Yndur\'ain and C. P. Yuan for 
useful comments. This work has been partially supported by Conacyt (M\'exico).



\begin{thebibliography}{0}    

\bibitem{orr}
I.~I.~Y.~Bigi, Y.~L.~Dokshitzer, V.~A.~Khoze, J.~H.~Kuhn and P.~M.~Zerwas,
  Phys.\ Lett.\  B {\bf 181}, 157 (1986)
\bibitem{orr-1}
L.~H.~Orr,  Phys.\ Rev.\  D {\bf 44}, 88 (1991).

\bibitem{willenbrock}
S. Willenbrock, Rev. Mod. Phys. {\bf 72}, 1141 (2000).
\bibitem{willenbrock-1}
P. C. Bhat, H. B. Prosper and S. S. Snyder, Int. J. Mod. Phys. A {\bf 13},
5113 (1998). 
\bibitem{willenbrock-2}
M.~Beneke {\it et al.}, arXiv:hep-ph/0003033.

\bibitem{tevatron}
See http://tevewwg.fnal.gov/top/.

\bibitem{fujii97}
K. Fujii, Nucl. Phys. Proc. Suppl. B {\bf 59}, 331 (1997).

\bibitem{pdg2006}
W.~M.~Yao {\it et al.}  [Particle Data Group],
  J.  Phys. G {\bf 33}, 1 (2006).

\bibitem{jk}
M. Jezabek and J. H. K\"uhn, Nucl. Phys. B {\bf 314}, 1 (1989).
\bibitem{jk-1}
M. Jezabek and J. H. K\"uhn, Phys. Rev. D {\bf 48}, R1910 (1993).

\bibitem{alpha_s2}
A. Czarnecki and K. Melnikov, Nucl. Phys. B {\bf 544}, 520 (1999).
\bibitem{alpha_s2-1}
K. G. Chetyrkin, R. Harlander, T. Seidensticker and M.
Steinhauser, Phys. Rev. D {\bf 60}, 114015 (1999).

\bibitem{alpha}
A. Denner and T. Sack, Nucl. Phys. B {\bf 358}, 46 (1991). 
\bibitem{alpha-1}
G. Eilam, R. R. Mendel, R. Migneron and A. Soni, Phys. Rev. Lett.
{\bf 66}, 3105 (1991).

\bibitem{renorm}
S. M. Oliveira, L. Bruecher, R. Santos and A. Barroso, Phys. Rev. D {\bf 64}, 
017301 (2001).

\bibitem{higgs}
B. Grzadkowski and W. Hollik, Nucl. Phys. B {\bf 384}, 101 (1992).
\bibitem{higgs-1}
A. Denner and A. Hoang, Nucl Phys. B {\bf 397}, 483 (1993).

\bibitem{Soni} 
D. Atwood, G. Eilam, A. Soni, R. R. Mendel and R. Migneron, Phys. Rev.
Lett. {\bf 70}, 1364 (1993).

\bibitem{Bernreuther}
W.~Bernreuther, A.~Brandenburg, Z.~G.~Si and P.~Uwer,
Nucl.\ Phys.\  B {\bf 690}, 81 (2004).
\bibitem{Bernreuther-1}
A.~Brandenburg, Z.~G.~Si and P.~Uwer,
Phys.\ Lett.\  B {\bf 539}, 235 (2002).
\bibitem{Bernreuther-2}
S.~Dittmaier and M.~Roth,
Nucl.\ Phys.\  B {\bf 642}, 307 (2002).

\bibitem{kz} 
See for example: N.~Kauer and D.~Zeppenfeld,
  Phys.\ Rev.\  D {\bf 65}, 014021 (2002).

\bibitem{muta}
T. Muta, R. Najima and S. Wakaizumi, Mod. Phys. Lett. A {\bf 1}, 203 (1986).

\bibitem{tbwz}
G. Altarelli, L. Conti and V. Lubicz, Phys. Lett. B {\bf 502}, 125 (2001).

\bibitem{wprop1}
M. Beuthe, R. Gonz\'alez Felipe, G. L\'opez Castro and J.
Pestieau, Nucl. Phys. B {\bf 498}, 55 (1997).
\bibitem{wprop1-1}
D. Atwood, G. Eilam,
R. Mendel, R. Migneron and A. Soni, Phys. Rev. D {\bf 49}, 289
(1994).

\bibitem{fls}
E. N. Argyres {\it et al.}, Phys. Lett. B {\bf 358}, 339 (1995).
\bibitem{fls-1}
U. Baur and D. Zeppenfeld, Phys. Rev. Lett. {\bf 75}, 1002 (1995).

\bibitem{Genaro}
G. L\'opez Castro and G. Toledo Sanchez,
  Phys.\ Rev.\  D {\bf 61}, 033007 (2000). 

\bibitem{fjy}
Private communications with F. J. Yndur\'ain and C. P. Yuan.

\bibitem{smith}
M. C. Smith and S. Willenbrock, Phys. Rev. D {\bf 54}, 6696 (1996)

\end{thebibliography}
\end{document}